\documentstyle[prl,aps,twocolumn]{revtex}

\newcommand{\eqn}{\begin{equation}}
\newcommand{\eqnend}{\end{equation}}

\newcommand{\mod}[1]{\mid #1 \mid}
\newcommand{\expval}[1]{\langle #1 \rangle}

\newcommand{\fppicture}{\begin{picture}(400,600)}

\tighten
\begin{document}
\draft
\twocolumn[
\hsize\textwidth\columnwidth\hsize\csname @twocolumnfalse\endcsname
\title { Diffusive and ballistic motion in superconducting
hybrid structures.}
\author{N.R. Claughton, R. Raimondi, and C.J. Lambert }
\address{School of Physics and Chemistry, Lancaster University, Lancaster,
LA1 4YB, England}
\date{\today}
\maketitle
\begin{abstract}
We examine transport
 properties of superconducting hybrid mesoscopic structures,
in both the  diffusive and  ballistic regimes.
For diffusive structures, analytic results
from quasi-classical theory
 are compared with predictions from numerical, multiple-scattering
calculations performed on small structures. For all structures, the two
methods yield comparable results and in some cases, quantitative
agreement is obtained. These results not only
demonstrate that quasi-classical
theory can yield the ensemble averaged conductance $<G>$
of small structures of
dimensions of order 10- 20 Fermi wavelengths, but also establish that
numerical scattering calculations on such small structures can
yield results for
$<G>$ which are
characteristic of much larger systems.
Having compared the two approaches,
we  extend the multiple-scattering analysis
to the ballistic
limit, where the
sample dimensions become smaller than the elastic
mean free path and demonstrate that the properties of certain Andreev
 interferometers
are unchanged in the clean limit.
\end{abstract}
]
\narrowtext

\section{Introduction}
\label{1}

When a normal metal (N) makes contact with a superconductor (S), classical
tunnelling theory predicts that as a consequence of the existence of a
superconducting energy gap, the low-temperature sub-gap conductance will
 vanish.
Andreev scattering \cite{an64} provides an alternative mechanism for charge
transport through such an N-S junction and for
an ideal interface \cite{btk82},
leads  to a zero-voltage conductance which is almost twice that of the normal
state. In the presence of an insulating layer (I), Andreev scattering
 becomes
less effective and in the absence of disorder, the sub-gap conductance of an
N-I-S junction is predicted to be depressed  compared with that of the normal
state. In contrast, at low enough temperatures, experiments on N-I-S junctions
\cite{ka91} reveal the existence of a zero-voltage conductance peak, with a
value comparable to the conductance in the normal state. This effect
is due to
the interplay between disorder-induced scattering and tunnelling and has been
observed in experiments  involving quantum wells with superconducting
 electrodes
\cite{cha92}, superconductor-normal metal microjunctions \cite{pe93}, and
superconductor-2DEG-superconductor structures \cite{ba94}. These have been
interpreted using a number of theoretical approaches, including
 quasi-classical
Green function techniques \cite{za90}-\cite{zai94},
  tunnelling Hamiltonian methods  \cite{he93},\cite{he94},
multiple-scattering techniques \cite{la91}-\cite{ma93},
 and random matrix theory \cite{be93},\cite{be94}.

Recently, following a number of
theoretical papers on disordered transport in the presence of two
superconducting contacts \cite{spi82}-\cite{la93},
several new experiments aimed at probing the phase-coherent nature of
Andreev scattering have been carried out. These involve a normal metal in
contact with superconductors $S_1$ and $S_2$, with order parameter phases
$\phi_1$, $\phi_2$, whose order parameter
phase difference $\phi=\phi_1-\phi_2$
can be controlled by external means.
For a diffusive system of size $L$, with diffusion coefficient $D$,
early theoretical work \cite{spi82},\cite{al87} predicted that
in the high temperature limit $T>T^*$, where
$k_BT^*=hD/L^2$,  the ensemble averaged conductance
$<G>$ should be a periodic function of $\phi$, with fundamental period $\pi$,
whereas in the low temperature regime $T<T^*$, it was predicted
\cite{la93},\cite{hui93}
that the
fundamental periodicity of $<G>$ would be $2\pi$. This prediction
of a $2\pi$  periodicity is a common feature of all recent
theories of $<G>$ \cite{he93},\cite{he94},\cite{za94},\cite{ta94}
and of recent  experiments \cite{pr94}-\cite{pe95}.

Despite much progress,
 many details of such Andreev interferometers
remain to be understood. In particular, there exists no general
theory of the amplitude of oscillation, the nature of the zero-phase
 extremum
and the harmonic content of the conductance-phase characteristic.
 Experiments on
various geometries, in different transport regimes
 have yielded amplitudes of
oscillation which differ by many orders of magnitude.
It has also been
questioned whether analytical theories that mainly
apply in the diffusive limit
can be used in the quasi-ballistic case \cite{di95}.

One difficulty in establishing a general theory is that
 most theoretical papers
are based on a single technique, with little detailed
 comparison with  the
results of other approaches. For example, distinct analytical
 theories exist for
the ballistic, diffusive and strongly disordered regimes, but
there is no
analytical theory capable of describing the cross-over between
 them. In
contrast, exact numerical solutions of the Bogoliubov - de Gennes
equation
\cite{lhr93} can easily cross from one regime to another, but are
 limited to
system sizes of order a  hundred Fermi wavelengths.

In this paper, we undertake a detailed comparison
 between an exact numerical,
multiple-scattering
technique and quasi-classical theory. The former was first used to
solve the Bogoliubov - de Gennes equation for disordered,
 one-dimensional
systems \cite{hl90} and soon after generalized to higher
 dimensions \cite{lh90},
while the latter,  developed in the context of nonequilibrium
superconductivity\cite{eil68}-\cite{lar77},
when supplemented by the appropriate boundary conditions
 \cite{za84},\cite{ku88},
has recently  yielded a variety of results
for ensemble averaged conductances  in the diffusive
limit\cite{za90}-\cite{zai94}.
  Once  agreement between the two
 approaches is obtained  in the
diffusive limit, we then  use the numerics to follow the
crossover  to the
ballistic regime, where the electron mean free path becomes
comparable with the
system size.

The comparison will be carried out by examining two canonical
 examples of
mesoscopic superconducting hybrid structures, namely an N-I-S
 junction and an
Andreev interferometer. In section II we briefly recall some
 results of the
quasi-classical Green function approach for these structures
 and in section III
highlight the main features of the numerical scattering approach.
 Whereas
a quasi-classical approach yields
 a one-dimensional theory whose results depend only on the
topology of the structure, the numerical calculations,
 in common with real
experiments, require the specification of a suitable geometry.
 In section IV we present
a detailed discussion of the
 geometry and  numerical parameters needed to reproduce the
 results of quasi-classical
theory. Having established agreement between the two
approaches  we then depart from the diffusive limit and
investigate the cross-over to ballistic transport, which
 is particularly
relevant to the experiments of reference \cite{di95}.

\section{Results from quasi-classical theory.}
\label{2}
In this section, we highlight some predictions of the
 quasi-classical approach
of references \cite{za90}-\cite{zai94}.
 These theories focus on the ensemble averaged
conductance $<G>$ and ignore the weak localisation
 contribution
discussed in \cite{spi82},\cite{al87}. The latter does not
scale with the system size and
in systems with a conductance much larger than $2e^2/h$, can
 be neglected.
The results of references  \cite{za90}-\cite{zai94} are obtained
 by solving the
following equation for the quasi-classical Green function
 $\hat g$ in the diffusive
limit
$D\partial_r(\hat g\partial_r\hat g)+\imath E[\hat\tau_z,\hat g]=0$,
 where $E$
is the quasi-particle excitation energy\cite{nota}.
  In what follows, we shall
consider only the $E=0$ limit of this equation, which
 applies to a diffusive system, whose length $L$ is
assumed to satisfy the inequalities   $L\ll \xi$,
 where $\xi^2 =D/\Delta_0 $
and
$\xi^2 =D/E $, in the superconducting and normal region
 respectively. Here $D$
is the diffusion coefficient, $\Delta_0$ the energy gap
and $E=eV$, where $V$ is
the applied voltage. Furthermore all temperatures
and voltages are assumed to be
much smaller than $\Delta_0$. For convenience in
 what follows, we shall also
restrict the analysis to zero temperature.

In
units of $2e^2/h$
 theory predicts\cite{na94},\cite{be94}
 that the total conductance $G$ of the structure
shown in figure \ref{g1}(a) may be computed from the equation
\begin{equation}
\label{eq:fish}
\frac{1}{G} = \frac{1}{G_{diff}} +  \frac{1}{G_{tun} \sin \theta},
\end{equation}
where $\theta$ is a solution of the transcendental equation
\begin{equation}
\label{eq:chips}
G_{diff} \, \theta = G_{tun} \, \cos \theta
\end{equation}
and $G_{tun}$, $G_{diff}$ are the respective conductances of the
tunnel junction and diffusive region in isolation.

There are two obvious limits to take. The first is where
$G_{tun}/G_{diff} \rightarrow 0$ for which
$\theta \approx G_{tun}/G_{diff} \approx \sin \theta$ and hence
\begin{equation}
\label{eq:cow}
\frac{1}{G}  \approx \frac{G_{diff}}{ G_{tun}^2}.
\end{equation}

The second limit is $G_{tun}/G_{diff} \rightarrow \infty$ in which
$\theta \rightarrow \pi / 2$. This  yields
\begin{equation}
\label{eq:bull}
\frac{1}{G} \approx \frac{ 1}{G_{diff}} + \frac{ 1}{G_{tun}}.
\end{equation}

Eq.(\ref{eq:cow}) has been also directly obtained in the
 tunneling Hamiltonian limit
\cite{he93}. Furthermore, as emphasized in Ref.\cite{be94},
the change of the power
in the dependence on the tunnel junction conductance $G_{tun}$,
as described by the equations
(\ref{eq:cow}) and (\ref{eq:bull}), reflects the combined effect
 of the Andreev scattering
at the N-S interface and the interference effects in the
mesoscopic phase coherent disordered
normal region. In the regime described by eq.(\ref{eq:bull})
and when $G_{tun}\ll 1$, the conductance of the
N-S junction appears to be the same as in the normal state and
the total resistance is obtained by simply adding up
in series the resistances of the diffusive region and tunnel junction.
Equations (\ref{eq:fish}) and (\ref{eq:chips})
 may be solved numerically to yield
the resistance $R_{tot}=1/G$ of the system as a
 function of $G_{diff}$ and
$G_{tun}$. The result of such an exercise for a
fixed value $G_{diff}=1.6$ and a
variable $G_{tun}$ is shown by the  curve (a) of
figure \ref{g4}. Also plotted
are the limits given by equations~(\ref{eq:cow})
and ~(\ref{eq:bull})
represented by the dashed lines (e) and (d).
The  curve (b) shows the results of
the numerical simulation described in section IV.

As a second example, quasi-classical
 theory predicts\cite{na94} that the conductance $G$ of the
structure depicted in figure \ref{f1}(a) may be
computed from the equation
\begin{equation}
\label{eq:fly}
G = \frac
{4 \, G_1^2 \, G_2^2 \, \cos^2 (\phi/2) }
{\left\{ G_1^2 + 4 \, G_2^2 \, \cos^2 (\phi/2) \right\}^{3/2}},
\end{equation}
where $G_1$ is the conductance of the tunnel junction (1),
$G_2$ is the
conductance of the tunnel junction (2) and $\phi$ is the
 phase difference
between the two superconductors.
In the limit $G_1 \gg G_2$, this simplifies to the expression
\begin{equation}
G = 4 \, \frac{G_2^2}{G_1} \cos^2 (\phi/2)
\end{equation}

whereas if $G_2 \gg G_1$ then
\begin{equation}
G = \frac{1}{2}\,\frac{G_1^2}{G_2} \,\frac{1}{\mod{\cos^2 (\phi/2)}}.
\end{equation}

Figure \ref{f3} shows four plots of equation~(\ref{eq:fly})
 for each combination
of $G_1 = 0.2, 2.0$ and $G_2 = 0.2, 2.0$. One can see that
 for $G_1 = G_2$ there
is a zero-phase minimum, as there is for $G_1 \ll G_2$.
 When $G_1 \gg G_2$
however, one finds a zero-phase maximum. In all cases
the conductance vanishes
when the phase difference between the superconductors is $\pi$.
 Corresponding
analytical results have been obtained for ballistic
 interferometers in
one\cite{nt91}\cite{t92} and two\cite{asrl95} dimensions.

\section{A scattering approach to transport
in mesoscopic superconductors.}
\label{3}
During the past six years \cite{lhr93}-\cite{lh90}
  we have
developed numerical codes capable of yielding exact
 solutions of the Bogoliubov
- de Gennes equation for disordered structures in
arbitrary dimensions.
Currently there are two independent sets of codes
available at Lancaster; one of
these is based on a transfer matrix approach and
the other is based on a
recursive Green's function method. Typically these
are used as independent
cross-checks and therefore there can be no doubt
about the accuracy of the
results obtained for a given structure.
 For a two-dimensional system of  width
less than a few hundred Fermi wavelengths,
or for a three-dimensional system of
width less than a few tens of Fermi
wavelengths, the problem of computing dc
transport properties of a phase-coherent
 system described by mean-field BCS
theory is therefore no longer an issue.
 Just as the appearance of pocket
calculators rendered approximate methods
for computing elementary functions
redundant, the existence of these codes
has, for several years, allowed
transport properties of small structures
to be calculated without further
approximation. For larger systems,
the key issue is how to extrapolate
the results of such calculations to
larger numbers of channels. By making
contact with quasi-classical theory,
the results which follow establish
that for many  systems, the ensemble
averaged conductance obtained from
small systems exhibits the essential
features of much larger structures.

The numerical codes evaluate  multi-channel
scattering formulae for
the dc
electrical conductance\cite{la91}\cite{lhr93} and more
recently  thermoelectric properties\cite{cl95}.
In what follows, we focus on hybrid structures
connected to normal external
reservoirs only\cite{note}. For a structure
connected to two such reservoirs,
the zero-temperature, zero-bias electrical
conductance can be written
\cite{la91},\cite{lhr93} (in units of $2e^2/h$),
\begin{equation}
G=T_0+T_a + {{2(R_a R'_a -T_a T'_a)}\over {R_a+R'_a+T_a+T'_a}}
\label{3.1}
\end{equation}

In this expression, $R_0, T_0$ ($R_a, T_a$) are the coefficients
for normal
(Andreev) reflection and transmission for zero-energy
quasi-particles from
reservoir $1$, while $R'_0, T'_0$ ($R'_a, T'_a$)
are the corresponding
coefficients for quasi-particles from reservoir $2$.
 If each of the external
leads connecting the reservoirs to the scatterer
 contains $N$ open channels,
these satisfy
 $R_0+T_0+R_a+T_a=R'_0+T'_0+R'_a+T'_a=N$ and $T_0+T_a=T'_0+T'_a.$
Furthermore, in the absence of a magnetic field,
 all reflection coefficients are
even functions of $\phi$, while the transmission
coefficients satisfy
$T'_0(\phi)=T_0(-\phi)$, $T'_a(\phi)=T_a(-\phi)$.
 Consequently on quite general
grounds, in the absence of a magnetic field, $G$
is predicted to be an even function of
$\phi$.

In the absence of quasi-particle transmission between
the two external probes,
equation (\ref{3.1}) reduces to
 \begin{equation} G^{-1}=(2R_a)^{-1}+(2R'_a)^{-1}
\label{3.2}, \end{equation} where $2R_a$  ($2R'_a$)
are left (right) boundary
conductances, introduced by Blonder, Tinkham and Klapwijk\cite{btk82}.
 The Lancaster codes yield all possible scattering
coefficients, but in what follows, to compare results
 with the theory
leading to eqs.(\ref{eq:fish}) and (\ref{eq:fly}), we
shall analyze structures with no transmission and restrict
 attention to $R_a$
only.

The numerical codes compute the scattering coefficients of
 a tight-binding
lattice, described by a
Bogoliubov - de Gennes (BdG) Hamiltonian of the form
\begin{equation}
{H}=\left(\matrix{{ H_0}&\Delta\cr\Delta^*&-{ H_0}^*}\right).
\end{equation}

If an index $n$ is used to label sites on the lattice and any internal
spin-degrees of freedom, then $ H_0$ is of the form
\begin{equation}
 { {H_0}} ={\Large \sum_{n}} \epsilon_{n}|n\rangle
 \langle n|  + {\Large \sum_{(n,m)}} V_{n,m}|n\rangle
 \langle m|.
\end{equation}

To model a given physical structure, it is necessary
to specify certain
phenomenological parameters which capture the
 essential physics. As an example,
in the absence of spin-orbit scattering, spin degrees
 of freedom can be ignored
and in the absence of a magnetic field, one chooses
 $V_{n,m}=-\gamma$ for
nearest neighbour  pairs $(n,m)$. If $(n,m)$ are
 not nearest neighbours, then
$V_{n,m}=0$. In a region free from disorder, the
 diagonal elements $\epsilon_n$
are set equal to a constant $\epsilon_0$, whereas
 in a disordered region,
$\epsilon_n$ is a random number uniformly
 distributed between $\epsilon_0 - W$
and $\epsilon_0 +W$. In the presence of
 spin singlet, local s-wave pairing,
$\Delta$ is a diagonal order parameter
 matrix with elements $\Delta_n$. In a
normal region, $\Delta_n=0$, whereas in a
 clean superconducting region,
$\vert\Delta_n\vert$ is set to a constant
 value $\Delta_0$. The phase of
$\Delta_n$ is chosen to equal a value
assigned to the superconducting region to
which site $n$ belongs. In what follows,
the energy scale will be fixed by
making the choice $\gamma=1$.

Of course, the above parameters are not directly
 accessible experimentally and
are not an explicit feature of quasi-classical
theory. Therefore when making
comparisons, some effort is needed to map one
analysis onto another. In
$d$-dimensions, for a clean system on a square
or cubic lattice, the chemical
potential relative to the band bottom is $\mu=\epsilon_0+2d\gamma$,
 the band
width is $4d\gamma$ and the effective mass for excitations near the
band bottom
is $m^*=\hbar^2/(2\gamma a^2)$, where $a$ is the lattice constant.
 A key parameter
in the problem is the dimensionless ratio $\bar\Delta=\Delta_0/\mu$,
which takes
a value $10^{-3}$ for conventional low $T_c$ superconductors such as
 Niobium,
but can be as large as $0.1$ for high-temperature superconductors,
 or for a 2DEG
in contact with a conventional superconductor.
 Andreev's approximation,  which
underpins many analytic theories,
including quasi-classical  and random matrix
descriptions, is valid only when this parameter is much less
 than unity.

Other parameters which are needed when making comparisons are
the Thouless
energy $E^*$, which for a  diffusive structure of width $M$,
length $L_{diff}$
and normal-state conductance $G$, is given by
$E^*=hD/L_{diff}^2=(h/2e^2)G/(n(0)L_{diff}M^{d-1})$, where
 $n(0)$ is the density
of states per site. A second parameter is the normal-state,
 elastic mean-free
path $l$, which for a diffusive sample connected to external
lead with $N$ open
channels, is given by $G=(2e^2/h)Nl/L_{diff}$.
 Within a numerical simulation on
a given geometry, once the model parameters $W$, $\gamma$, $\epsilon_0$ and
$\Delta_0$ are chosen, the parameters $G$, $l$, $n(0)$ and $E^*$ are computed
explicitly.

\section{Numerical results for a N-I-S structure. }
\label{4}

In this section, we present a comparison between the
predictions of quasi-classical
theory and the above numerical scattering approach,
for an N-I-S structure.
Our aim is to highlight the steps required to obtain
a suitable choice of
parameters, from which a meaningful comparison can be made.
In
the literature, numerical results in two-dimensions
have been obtained by  first
solving for the scattering matrix of a normal
diffusive structure, with or
without a tunnel junction and then employing
 Andreev's approximation to model
the Andreev scattering induced by a
nearby superconductor\cite{ma93}. As noted
above, this approximation requires
 that $\Delta_0$ be small compared with the
Fermi energy and that there be no disorder in the
superconductor. Furthermore
for a clean system, Andreev's approximation
can yield incorrect results, because
even at a clean N-S interface, the approximation
 breaks down \cite{cl95} when
scattering channels are almost closed.
 For these reasons a comparison with an
exact solution of the Bogoliubov - de Gennes
 equation allows one to examine
changes occurring away from the Andreev limit,
in a region of parameter space
which is more relevant to high-temperature superconductors.

The system to be examined is shown in figure \ref{g1}(b)
 and consists of a
disordered region in contact with a tunnel junction,
which is in turn adjacent
to a superconducting probe. The simulated structure
consists of a
two-dimensional tight-binding lattice of width $M$
sites. The disordered region
is of length $L_{diff}$ sites, the tunnel
 junction is $L_{tun}$ sites long and
the superconductor has a length $L_{sup}$.
In units of $2e^2/h$, the conductance
of a particular realisation of the structure
will be denoted $G$ and the
ensemble-averaged conductance will be written
 $\expval {G }$. In order to make a
comparison with quasi-classical theory,
 it is necessary that the properties of the two
resistive components and the superconducting probe
 be compatible with the
assumptions made by the theory.

To identify a suitable choice of parameters,
consider first a normal diffusive
portion of length $L_{diff}$ and width $M$,
connected to crystalline leads. In
units of $2e^2/h$, the conductance of
 a particular realisation of this
structure  will be
denoted $G_{diff}$ and the ensemble-averaged
conductance will be written
$\expval {G_{diff} }$. The conductance of a
diffusive material is inversely
proportional to its  length and therefore a
plot of $\expval {G_{diff} }
L_{diff}$ as a function of $ L_{diff}$ will
exhibit a plateau in the diffusive
regime, with a mean free path given by
$l=\expval{G_{diff}}L_{diff}/N$.

For a sample of width $M=10$, figure \ref{g2} shows a plot of
$\expval{G_{diff}}L_{diff}/N$ versus $L_{diff}$.
 This structure has periodic
boundary conditions  in the  direction tranverse
 to the current flow and the
choice $\epsilon_0 = 0.2$ was made, which yields $N=9$.
 Results are shown for a
disorder of $W=1$. The length $L_{diff}$ of the disordered
 region was
incremented in steps of two sites from  $L_{diff}=2$ to $40$.
 For each value of
$L_{diff}$, 2000 realizations of disorder were chosen
 and the conductance
$G_{diff}$ computed  for each. Then the ensemble average
 $\expval {G_{diff} }$
and the standard deviation $\delta {G_{diff} }$ were calculated.

Figure \ref{g2} shows that in the interval $20 < L_{diff} < 40$
the system
exhibits diffusive behaviour, with a mean free path of $l \approx 4.6$.
 For
smaller values of $L_{diff}$, the system is in the ballistic regime and
 for
larger values, the onset of localisation causes the curve to fall.
 A diffusive
system is one for which $l \ll L_{diff}$ and $l \ll M$. Furthermore if weak
localization corrections are to be neglected,
we require $N l \gg L_{diff}$. For
these reasons a judicious choice of length
 yielding a diffusive system whilst
minimizing CPU time is  $L_{diff} = 25$.
 Such a system has an average
conductance $\expval {G_{diff} } = 1.6$

To compare with quasi-classical
 theory, a knowledge of the conductance of the isolated
tunnel junction $G_{tun}$ as a function of the barrier
 height $\epsilon_b$ is
also required. In what follows, we consider a clean
 tunnel junction of
dimensions $M=10$ and $L_{tun}=1$, obtained by setting
the diagonal elements
$\epsilon_n$ of all barrier-sites $n$ equal to
$\epsilon_0 +\epsilon_b$.   For
an isolated barrier connected to crystalline
leads of width 10, figure \ref{g3}
shows a plot of $G_{tun}$ as a function of $\epsilon_b$.
 To obtain this plot,
the conductance $G_{tun}$ is computed for 1000 successive
 values of the barrier
height $\epsilon_b$  in the range $ 0.0 < \epsilon_b < 10.0$.
 This choice of
barrier heights yields a spread of barrier conductances in the
convenient range
$ 0.0 < G_{tun} < 9.0$.

Finally, before a comparison with  theory can be made,
the properties of
the superconductor must be chosen such that there be no
 transmission through the
superconducting region and that Andreev's approximation
 of neglecting normal
reflection at the N-S interface is valid. To avoid quasi-particle
transmission,
it is necessary to choose $L_{sup} > \xi$, where $\xi=\mu/\Delta_0$ and to
minimise normal reflection it is necessary that $\xi >> 1$. For  $\Delta_0 =
0.05$ and $\epsilon_0=0.2$ the superconducting
coherence length is $\xi =76$ and
it is found that transmission is negligible for $L_{sup} > 100$.
 For the above
choice of parameters, one finds for the normal and Andreev
 reflection and
transmission coefficients:  $R_0 = 0.06485$,
 $R_a = 8.84762$,  $T_0 = 0.08748$,
and  $T_a = 0.00005$.

It should be noted that the condition $\xi >>1$
is not sufficient to completely
exclude normal reflection at an N-S interface.
It is also necessary that
$\epsilon_0$ be chosen such that the number of
 open channels in the external
leads is not sensitive to small changes in $\epsilon_0$.
This feature is
illustrated in figure \ref{steps}, which shows as a
function of $\epsilon_0$,
the conductance of a clean superconducting region of width $M=50$
 and length
$L_{sup}=5$, attached to crystalline normal leads.
Results are shown for 5
values of $\Delta_0$: $\Delta_0=0, 10^{-2}, 10^{-1}, 0.3$ and $0.5$. For
$\Delta_0=0$, the conductance is equal to the number of open channels and
changes by unity whenever an external quasi-particle channel closes.
 At these
values of $\epsilon_0$, switching on an infinitesimal $\Delta_0$
 causes the
conductance to decrease by unity. As shown in the figure,
finite values of
$\Delta_0$ smear the conductance steps and suppress
 the conductance. Both of
these features lie outside Andreev's approximation.
To achieve compatibility
with the assumptions of circuit theory, the choice
$\epsilon_0=0.2$ was made,
which places a system of width 10 between two
conductance steps and avoids the
above sensitivity to changes in $\Delta_0$.

Having identified a choice of parameters which is compatible
 with quasi-classical
theory, numerical results for the combined structure
 of figure \ref{g1}b can now
be obtained. To summarize, this structure has the
 following properties: width
$M=10$,  number of open channels $N=9$, band filling factor $\epsilon_0 =
0.2$ leading to a chemival potential  $\mu = 3.8$,
 length of tunnel junction $L_{tun}=1$, barrier heights $
0.0 <\epsilon_b < 10.0$, barrier conductances
 $ 0.0 < G_{tun} < 9.0$, length of
diffusive region $L_{diff}=25$, diffusive disorder width $W=1$,
 conductance of
diffusive region $\expval {G_{diff}} = 1.6$,  length of
 superconductor
$L_{sup}=100$,  superconducting coherence length $\xi=76$, superconducting
order parameter $\Delta_0=0.05$, elastic mean free path $l \approx 4.5$.

First consider the case of no barrier, where $\epsilon_b=0$.
 In this case,
quasi-classical theory insists that the conductance of a
 diffusive region in contact
with a superconductor should be identical with
 the normal-state conductance of
the diffusive region. Figure \ref{nsplot} shows
 plots of the mean conductance
$\expval{G}=N-R_0(\Delta_0) + R_a(\Delta_0)$ as a
 function of $\Delta_0$, for
disordered regions of four different lengths. In
the normal state
$(\Delta_0=0)$ $\expval{G}$ reduces to $T_0(0)=N-R_0(0)$
 and in the presence of
a sufficiently-long superconductor, to the BTK
 conductance $2R_a(\Delta_0)$. The
left insert shows the quantity
 $<g>= \expval{N-R_0(\Delta_0)+ R_a(\Delta_0)}
/\expval{T_0(0)}$ (ie the  conductance divide
 by the normal state conductance).
The right insert shows the root mean square
 deviation
$\sigma=\expval{({G(\Delta_0)}-\expval{G(\Delta_0)})^2}^{1/2}$.
 These show that
in the ballistic limit $L_{diff}=0$, the conductance rises
 to a value almost
double that of the normal state, before decreasing with
 increasing $\Delta_0$.
In contrast, the mean conductance of a diffusive normal
 region is relatively
insensitive to the onset of superconductivity, with the
 largest relative change
corresponding to the largest value of $L_{diff}$,
 ( ie the smallest value of the
normal state conductance). It should be noted
 however that even though the mean
is insensitive to $\Delta_0$, the fact that
the rms deviation $\sigma$ is
non-zero reveals that for individual samples,
large changes of arbitrary sign
can occur.

Having examined a diffusive conductor with no barrier,
 we now turn to the case
of finite $\epsilon_b$. Curve (b) of figure \ref{g4}
shows numerical results in
the presence of a tunnel barrier. For 50
equally-spaced barrier heights in the
range   $ 0.0 <\epsilon_b < 10.0$, 500 realizations of
 disorder in the diffusive
region were selected and the total conductance $G$
computed for each
realisation. The ensemble-averaged conductance
$ \expval{G}$ was then calculated
and finally the total resistance  $\expval{R_{tot}} = 1 / \expval{G}$
plotted
against the computed conductance ${G_{tun}}$ of the isolated
 tunnel junction.
Since the average conductance of the diffusive region $\expval{G_{diff}}$ is
also known, equation~(\ref{eq:fish}) can be evaluated to yield
the corresponding
analytical result, curve (a) of figure \ref{g4}.

This demonstrates that in the range of validity of quasi-classical theory,
quantitative agreement with the numerical scattering approach is obtained.

\section{Numerical results for Andreev Interferometers.  }

Having examined a simple N-I-S structure, we now compare numerical results
 for
the tight-binding structure of figure \ref{f1}b, with the predictions of
quasi-classical
theory for the one-dimensional system of figure \ref{f1}(a). The latter
comprises a tunnel junction connected by diffusive 1-D wires to a fork.
 Each of
the two arms of the fork is a diffusive wire, connected via tunnel
junctions to
infinitely long superconductors. The conductance of the diffusive wires is
assumed to be much greater than that of the tunnel junctions.

The two-dimensional tight-binding realisation of this structure is shown in
figure \ref{f1}(b), which consists of a tunnel junction (1) lying next to a
diffusive region which is in turn adjacent to two superconductors. The
superconductors are separated from each other by an insulating layer and from
the diffusive region by two identical tunnel junctions (2).
 The superconductors
$i=1,2$ have order parameter phases $\phi_i$, but are identical in every other
respect. In order that they may successfully  represent superconducting probes
of infinite length, they are chosen in such a way that quasiparticle
transmission through them is negligible.

The diffusive region is of length $L_{diff}$ sites, each tunnel junction is
$L_{tun}$ sites long and the superconductors have a length $L_{sup}$.
 To model a
superconducting reservoir,  $L_{sup}$ is again chosen sufficiently
large such
that there is negligible transmission through the superconductor.
The system
width  and the width of both the diffusive region and the
 tunnel junction (1) is
$M$ sites. On the right of the diffusive region, the three
 insulating layers are
each one site thick and therefore the superconductors are each of width
$M^\prime$ (where $2 M^\prime = M -3$).

In the simulation, the conductances $G_1$ and $G_2$ of the
tunnel junctions are
fixed at values which replicate the three situations of
 figure \ref{f3}, namely
$G_1 = G_2$, $G_1 \ll G_2$ and $G_1 \gg G_2$ to enable
 comparisons to be made
with the analytic results. In each case, the phase difference between
 the two
superconductors is varied and the total conductance $G$ plotted
as a function of
phase for a particular realisation of disorder in the diffusive
 region.
Ensemble-averaging over many disorder realisations yields the
 conductance
$\expval G$ which is independent of the microscopic configuration of
 the system
and may be usefully compared with the results of equation~(\ref{eq:fly})
(see
below) and figure \ref{f3}.

In what follows, we examine a sample of width $M=15$, with a diagonal
 matrix
element $\epsilon_0 = 0.2$ and periodic boundary conditions, for which
the
number of open channels is $N=13$. The disorder is chosen to be $W=1$
and again
from a graph of the form of figure 5, we obtain a mean free path of
$l \approx
4.9$. In most cases of experimental interest, the conductance of
 the diffusive
`wires' may be considered to be much greater than that of the
 tunnel junctions.
In order to take into account this situation in our numerical
 simulations, one
has to put some restrictions on the length of the diffusive region,
 since the
conductance decreases with length. A compromise must therefore
 be found between
the desire to increase the length into the diffusive regime
 and the wish to
decrease it in order to maintain a high conductance.
 In what follows a choice
$L_{diff} = 10$ is made, for which $\expval {G_{diff} } \approx 5.1$.

To create a tunnel barrier of length $L_{tun}=1$ and width $M$,
 all diagonal
elements $\epsilon_i$ of sites within the barrier were set to
 a value
$\epsilon_i = \epsilon_0 +\epsilon_b$. For an isolated tunnel
 junction (1) of
width $M=15$, the values $\epsilon_b = 3.52, \, 12.27$ yield
 respectively the
conductances $G_1 = 2.0, \,0.2$ and for an isolated tunnel
 junction (2) of width
$M'=6$, the values $\epsilon_b = 1.87, \, 7.75$ yield
conductances $G_2 = 2.0,
\,0.2$. These  values of $\epsilon_b$ were used in the
simulations of figure 9.
As in the previous section, the choice $\epsilon_0=0.2$,
 $\Delta_0 = 0.05$,
$L_{sup} = 100$ was made. For the superconductor alone,
 connected to crystalline
normal leads of width $M'=6$, the values of the normal
 and Andreev reflection
and transmission coefficients were found to be: $R_0 = 0.02155$,  $R_a =
4.92905$,  $T_0 = 0.04937$,  $T_a = 0.00003$. For such a structure,
 there are 5
open channels and as a consequence, the sum of these four
 coefficients is 5.
Finally, in order to model an insulating barrier,
the diagonal matrix elements
$\epsilon_i$ referring to a site $i$ on the barrier
 between the superconductors,
were each set to the large number $\epsilon_i = \epsilon_0 + 50$.

By combining the above components to yield the complete
structure of figure
\ref{f1}(b), one obtains the structure to be
 analysed numerically, whose
parameters are as follows: total width $M=15$,
 superconductor width
$M^\prime=6$,  number of open channels in
 normal lead $N=13$,  band filling
$\epsilon_0 = 0.2$, chemical potential $\mu = 3.8$, length of tunnel
 junctions
$L_{tun}=1$,  length of diffusive region $L_{diff}=10$,
  disorder
width $W=1.0$, conductance of diffusive region
 $\expval {G_{diff}} = 5.1$,
length of superconductor $L_{sup}=100$,
 superconducting coherence length
$\xi=76$ , superconducting order parameter $\Delta_0=0.05$.

To carry out the simulation, the conductances $G_1$ of tunnel
 junction (1)  and
$G_2$ of tunnel junction (2) were fixed and a particular
 realisation $\{
\epsilon_i \}$ of disorder in the diffusive region was
selected. Then the phase
difference $\phi = \phi_1 - \phi_2$ between superconductors $1$
 and $2$ was
varied from zero to $2 \pi$. This was done by fixing $\phi_1 = 0$
and choosing
$50$ evenly spaced values of $\phi_2$. For each value of $\phi$,
 $200$ different
diffusive regions were obtained and the conductance $G$ of the
whole system
computed for each. The ensemble-averaged conductance
$\expval {G }$ was then
calculated. The graphs of figure \ref{f2}
 show plots of $\expval {G }$ as a
function of the phase difference $\phi$ for the following
four combinations of
$G_1$ and $G_2$. They are  (a) $G_1 = 0.2$  $G_2 = 0.2$,  (b) $G_1 = 2.0$
  $G_2
= 0.2$, (c) $G_1 = 0.2$  $G_2 =2.0 $, (d) $G_1 = 2.0$  $G_2 = 2.0$.

Apart from the different vertical scales, the numerical results
of figure
\ref{f2} and the analytic results of figure \ref{f3} share
 many qualitative
features and also exhibit some interesting differences. Figure
 \ref{f2}(d) is
comparable with \ref{f3}(d); each exhibits a zero-phase minimum
and a further
minimum at $\phi = \pi$. Similarly \ref{f2}(b)    is comparable with
\ref{f3}(b); each exhibits a zero-phase maximum, with a minimum
at $\phi = \pi$.
The remaining curves compare less favourably. Whereas the
 analytic results of
figures \ref{f3}(a) and \ref{f3}(d) are necessarily identical,
 there is no such
restriction on the numerics and as shown in figure \ref{f2}(a),
decreasing the
conductances $G_1$ and $G_2$ can produce qualitative changes.
As a consequence,
figure \ref{f2}(a) possesses a zero-phase maximum,
 whereas figure \ref{f3}(a)
possesses a zero-phase minimum.

Figures \ref{f3}(c) and \ref{f2}(c) also reveal some differences.
 Each possesses
a zero-phase minimum, but at $\phi=\pi$, where the analytic result
 vanishes, the
numerical result is almost maximal. The inset in figure \ref{f2}(c)
 shows a
`blowing up' of the region   $\pi- 0.15 \leq \phi \leq \pi+ 0.15$,
 with $G_1 =
0.2$ and $G_2 =2.0 $.  The inset shows three curves, obtained by
 averaging over
different numbers of  disordered samples,
 namely 200,  1000 and 2000
realisations of the disorder.
These demonstrate that in contrast with eq.(\ref{eq:fly}),
 the numerical results \ref{f2}(c) possess a shallow,
local minimum at $\phi=\pi$.

Finally, we end this discussion by noting that for systems
 with a small number
of open channels, the behaviour of an individual sample
can be very different
from that of the mean. For each of the four cases (a) to (d),
 figure \ref{f5}
shows each of the 200 plots of conductance $G$ from which
 the ensemble averages
of figure \ref{f2} were calculated. Apart from the case  \ref{f5}(b),
 where
individual members of the ensemble behave in the same manner
 as the ensemble
average, the nature of the extrema at $\phi=0, \pi$ depends
on the miscroscopic
realisation of the disorder.
We also note that by changing the dimensions of the sample, one can
change the details of figure 9, but not the qualitative shape.
For example by
increasing the length $L_{diff}$ from 10 to 15, the local minimum
at $\phi=\pi$ in figure 9c becomes  more pronounced, but further
increasing $L_{diff}$ to 20 causes the minimum to become more shallow.

Having compared  the quasi-classical theory of references
 \cite{za90}-\cite{zai94}
 with the numerical scattering approach in the
diffusive limit, we now  examine the cross-over
 to the ballistic regime, where
the former is inapplicable. We focus attention
on the interferometer of figure
3(b) and examine the change in behaviour as the
length $L_{diff}$ of the
diffusive region becomes smaller than the elastic scattering
length $l$. Apart
from the change in $L_{diff}$ all other parameters are fixed
 to the values used
in figure 9.
Figure 11 shows results for a
diffusive region of length $L_{diff}=5$ and figure 12 for a length $L_{diff}
=1$.
Remarkably, apart from the overall increase in the conductance, the
qualitative shape of the curves is unchanged, despite the fact that
the restrictions on quasi-classical theory are violated by
these structures.

\section{Discussion}
\label{6}
In this paper, for the first time, a detailed comparison
 between quasi-classical
theory and numerical multiple-scattering calculations has
 been carried out. To ensure that the
simulated structures fall within the parameter range where
 the approximations of
quasi-classical Green function methods hold,
 we have painstakingly examined each component of a given
 structure. For
the N-I-S structures of figure 1, figure 2 shows that there is
 quantitative
agreement between the two methods. For the interferometers of figure 3,
 there is
broad qualitative agreement, although as shown in figures \ref{f2}
and \ref{f3},
some interesting differences are present. The theory of references
\cite{za90}-\cite{zai94} is a
quasi-one-dimensional theory and for the symmetric structures of figure 3,
necessarily predicts a vanishing conductance at $\phi=\pi$. This symmetry
is not
present at a microscopic level and therefore there is no such restriction
on the
results from an exact solution of the Bogoliubov - de Gennes equation.
 Figures
\ref{f2}(a,b,d) suggest that for certain structures, this microscopic
symmetry-breaking may be unimportant, but for other strengths of the tunnel
barriers, figure \ref{f2}(c) suggests that this artifact of quasi-classical
theory will
not be observed experimentally.
In the clean limit, figures 11 and 12 show that although the overall
conductance is increased, the qualitative shape of the
 conductance-phase curves is unchanged.

The above results demonstrate that quasi-classical theory yields the
correct shape for the ensemble-averaged conductance even down to
extremely small system sizes and that results obtained for dirty
systems can be applicable, in some cases, even in the clean limit.
They also demonstrate that even without attempting a systematic
extrapolation to a large number of channels,
numerical multiple-scattering calculations
on small structures,
can yield results for ensemble-averaged properties of much larger systems.

The results of figure \ref{f7} demonstrate that the essential properties
of these interferometers are unchanged in the clean limit and therefore as
already noted in \cite{asrl95} disorder is not a necessary feature of
large amplitude Andreev interferometers.

\acknowledgements
We would like to thank  A.F. Volkov, A.D. Zaikin and M.Leadbeater
for extended discussions. Financial support from the EPSRC, the MOD,
the Institute for Scientific Interchange  and
NATO is also gratefully acknowledged.

\narrowtext

\begin{figure}
\caption{(a) Schematic picture
of an N-I-S junction for analysis using circuit theory.
(b) Picture
of the tight-binding lattice used to model a two-dimensional
 N-I-S structure.}
\label{g1}
\end{figure}

\begin{figure}
\caption{Conductance of the N-I-S structures of figure 1,
 as a function of the
conductance $G_{tun}$ of the tunnel junction.
 The various curves
refer to:
a) analytic theory; b) numerical simulation;
c) resistance of the diffusive region $<R_{diff}>=0.63=<G_{diff}>^{-1}$;
d) asymptotics at high transparency $G_{tun}\gg <G_{diff}>$;
 e) asymptotics at low
transparency  $G_{tun}\ll <G_{diff}>$.}
\label{g4}
\end{figure}

\begin{figure}
\caption{  (a) Schematic picture of an interferometer
analyzed using circuit
theory. (b) Picture an interferometer formed from a two-dimensional
tight-binding lattice.}
\label{f1}
\end{figure}

\begin{figure}
\caption{Analytic results for the conductance versus phase
in the diffusive limit: a) $G_1 =0.2$ and $G_2=0.2$.
 b) $G_1 =2.0$ and $G_2=0.2$.
 c) $G_1 =0.2$ and $G_2=2.0$.
 d) $G_1 =2.0$ and $G_2=2.0$.}
\label{f3}
\end{figure}

\begin{figure}
\caption{$<G_{diff}>L_{diff}/N$ versus $L_{diff}$.
 The plateau region signifies
the diffusive regime. Here the number of open channels is $N=9$.
In the inset standard deviation is also shown.}
\label{g2}
\end{figure}

\begin{figure}
\caption{ Conductance $G_{tun}$ of the tunnel
junction as a function of the
barrier height $\epsilon_b$.}
\label{g3}
\end{figure}

\begin{figure}
\caption{ The conductance $G(\Delta_0)$ of a
clean superconducting region
of length $L_{sup}=5$, width $M=50$, plotted
as a function of the site
energy $\epsilon_0$, for 5 different values of $\Delta_0$.
}
\label{steps}
\end{figure}

\begin{figure}
\caption{Plots of the mean conductance
 $\expval{G(\Delta_0)}=N-R_0 + R_a$ as a
function of $\Delta_0$, for a diffusive
 conductor of width $M=10$ and four
different lengths. The left insert shows
 the  conductance
$<g>=\expval{G(\Delta_0) }/\expval{G(0)}$, scaled by the normal state
conductance. The right insert shows the rms deviation
$\sigma=\expval{({G(\Delta_0)}-\expval{G(\Delta_0)})^2}^{1/2}$.
}
\label{nsplot}
\end{figure}

\begin{figure}
\caption{Numerical results for the
conductance versus phase for
 $L_{diff}=10$:
 a) $G_1 =0.2$ and $G_2=0.2$.
 b) $G_1 =2.0$ and $G_2=0.2$.
 c) $G_1 =0.2$ and $G_2=2.0$.
 d) $G_1 =2.0$ and $G_2=2.0$}
\label{f2}
\end{figure}

\begin{figure}
\caption{Numerical results for the
conductance versus phase for individual realizations of the
 disorder :
 a) $G_1 =0.2$ and $G_2=0.2$.
 b) $G_1 =2.0$ and $G_2=0.2$.
 c) $G_1 =0.2$ and $G_2=2.0$.
 d) $G_1 =2.0$ and $G_2=2.0$.}
\label{f5}
\end{figure}

\begin{figure}
\caption{Numerical results for the
conductance versus phase for  $L_{diff}=5$:
 a) $G_1 =0.2$ and $G_2=0.2$.
 b) $G_1 =2.0$ and $G_2=0.2$.
 c) $G_1 =0.2$ and $G_2=2.0$.
 d) $G_1 =2.0$ and $G_2=2.0$.}
\label{f6}
\end{figure}

\begin{figure}
\caption{Numerical results for the
conductance versus phase for  $L_{diff}=1$:
 a) $G_1 =0.2$ and $G_2=0.2$.
 b) $G_1 =2.0$ and $G_2=0.2$.
 c) $G_1 =0.2$ and $G_2=2.0$.
 d) $G_1 =2.0$ and $G_2=2.0$.}
\label{f7}
\end{figure}


\begin{thebibliography}{99}
\bibitem{note} These codes are available on request.
 A version is also
available, in which one of the external reservoirs is a
 superconductor.
\end{thebibliography}
\end{document}